\documentclass
[preprintnumbers,superscriptaddress,unsortedaddress,onecolumn,11pt]{revtex4}%
\usepackage{amssymb}
\usepackage{amsmath}
\usepackage{graphicx}
\usepackage{dcolumn}
\usepackage{bm}
\usepackage{amsfonts}%
\providecommand{\U}[1]{\protect\rule{.1in}{.1in}}
\begin{document}
\title{Intermode Dephasing in a Superconducting Stripline Resonator - Supplementary Information}
\author{Oren Suchoi}
\affiliation{Department of Electrical Engineering, Technion, Haifa 32000 Israel}
\author{Baleegh Abdo}
\affiliation{Department of Electrical Engineering, Technion, Haifa 32000 Israel}
\author{Eran Segev}
\affiliation{Department of Electrical Engineering, Technion, Haifa 32000 Israel}
\author{Oleg Shtempluck}
\affiliation{Department of Electrical Engineering, Technion, Haifa 32000 Israel}
\author{M. P. Blencowe}
\affiliation{Department of Physics and Astronomy, Dartmouth College, Hanover, New Hampshire
03755, USA}
\author{Eyal Buks}
\affiliation{Department of Electrical Engineering, Technion, Haifa 32000 Israel}
\date{\today }

\begin{abstract}

\end{abstract}

\maketitle

This paper contains suplementary information for \cite{Suchoi_3133}. The
supplementary information is devoted to three main issues. In section I we
describe the fabrication process; in section II we present the derivation of
the Hamiltonian of the system and provide a more detailed discussion about the
properties of the microbridges; in section III the hysteretic response of the
resonator and the effect of heating are discussed.

\section{Fabrication processes}

The fabrication process starts with a high resistivity Si substrate coated
with SiN layers of thickness $100%
\operatorname{nm}%
$ on both sides. A $150%
\operatorname{nm}%
$ thick Nb layer is deposited on the wafer using magnetron DC sputtering.
Then, e-beam lithography and a subsequent liftoff process are employed to
pattern an Al mask, which defines the SSR and the SQUID leads. The device is
then etched using electron cyclotron resonance system with CF$_{4}$ plasma.
The nanobridges are fabricated using FEI Strata 400 Focus Ion Beam (FIB)
system \cite{Hao_192507,Hao_392,Bell_630, Datesman_928, Troeman_2152} at
accelerating voltage of $30%
\operatorname{kV}%
$ and Ga ions current of $9.7$ pA. The outer dimensions of the bridges are
about $150\times50%
\operatorname{nm}%
.$ However, the actual dimensions of the weak-links are smaller, since the
bombarding Ga ions penetrate into the Nb layer, and consequently, suppress
superconductivity over a depth estimated between $30%
\operatorname{nm}%
$ to $50%
\operatorname{nm}%
$ \cite{Troeman_2152, Datesman_3524}.

\section{Detailed derivation of the effective Hamiltonian}

The effective Hamiltonian of the closed system comprising the SSR and the
SQUID \cite{Clark_3042, Blencowe_014511} is found using the same method that
was previously employed in Refs. \cite{Blencowe_014511,Nation_104516}. Here
however, we relax the assumption that the self inductance of the SQUID loop is
small, and also the assumption that both junctions have the same critical
currents. On the other hand, we assume that the inductance of the SQUID, which
is denoted as $L_{\mathrm{S}}$, is much smaller than the total inductance of
the stripline $L_{\mathrm{T}}l_{\mathrm{T}}$. This assumption can be justified
by considering the fact that the measured angular resonance frequencies
$\omega_{n}$ of the first 3 modes ($n\in\left\{  1,2,3\right\}  $) for all
values of $\Phi_{x}$ (see Figs. 2 and 3 in \cite{Suchoi_3133}) are very close
to the values expected from a uniform resonator having length $l_{\mathrm{T}}%
$, namely $n\omega_{\mathrm{T}}$, where $\omega_{\mathrm{T}}=\pi
/l_{\mathrm{T}}\sqrt{L_{\mathrm{T}}C_{\mathrm{T}}}$. Moreover, the normalized
flux-induced shift $\Delta\omega_{n}/n\omega_{\mathrm{T}}$ in the angular
resonance frequency of the first 3 modes is quite small and never exceeds
$10^{-3}$. Both observations indicate that the ratio $L_{\mathrm{S}%
}/L_{\mathrm{T}}l_{\mathrm{T}}$ can indeed be considered as a small parameter.

The resultant Hamiltonian of the closed system is given by $\mathcal{H}%
=\mathcal{H}_{\mathrm{SSR}}+\mathcal{H}_{\mathrm{S}}\left(  I\right)  $, where
$\mathcal{H}_{\mathrm{SSR}}$ is the SSR Hamiltonian and where $\mathcal{H}%
_{\mathrm{S}}\left(  I\right)  $ is the SQUID Hamiltonian, which depends on
the current $I$ at the center of the SSR, namely, the current flowing through
the SQUID. In terms of annihilation ($A_{1}$ and $A_{3}$) and creation
($A_{1}^{\dag}$ and $A_{3}^{\dag}$) operators for the first and third modes of
the SSR respectively, the Hamiltonian $\mathcal{H}_{\mathrm{SSR}}$ can be
expressed as
\begin{equation}
\mathcal{H}_{\mathrm{SSR}}=\hbar\omega_{\mathrm{T}}\left(  N_{1}%
+3N_{3}\right)  +V_{\mathrm{in}}\;,
\end{equation}
where $N_{1}=A_{1}^{\dag}A_{1}$ and $N_{3}=A_{3}^{\dag}A_{3}$ are number
operators,%
\begin{equation}
V_{\mathrm{in}}=\hbar\sqrt{2\gamma_{\mathrm{f}1}}b_{1}^{\mathrm{in}}\left(
e^{-i\omega_{\mathrm{p}}t}A_{1}+e^{i\omega_{\mathrm{p}}t}A_{1}^{\dag}\right)
\end{equation}
represents the external driving, $\gamma_{\mathrm{f}1}$ is the coupling
constant between the 1st mode and the feedline, $b_{1}^{\mathrm{in}}$ is the
amplitude of the driving pump tone, which is injected into the feedline to
excite the first mode, and $\omega_{\mathrm{p}}$ is its angular frequency.

\subsection{The kinetic inductance of the nanobridges}

The Hamiltonian for the SQUID depends on the properties of the nanobridges.
Due to the Ga ions implanted in the outer layer of the Niobium during the FIB
process and the consequent suppression of superconductivity in that layer
\cite{Troeman_2152, Datesman_3524}, the weak links are treated as variable
thickness nanobridges. The behavior of such a nanobridge is strongly dependent
on the ratio $l/\xi$
\cite{Granata_275501,Hasselbach_4432,Hasselbach_140,Baratoff_1096,Likharev_101,Likharev_950,
Gumann_064529, Podd_134501}, where $l$ is the bridge length and $\xi$ is the
coherence length of the Cooper pairs. The coherence length $\xi$ depends also
on the temperature of the bridge. In the dirty limit $\xi$ is given by
$\xi(T)=0.852\sqrt{\xi_{0}l_{f}\left(  T_{C}/T-1\right)  ^{-1}}$
\cite{Likharev_101}, where $\xi_{0}$ is the size of the cooper pair and
$l_{f}$ is the mean free path\cite{Pronin_14416,Maxfield_A1515}. The
current-phase relation (CPR) of the bridges is periodic with respect to the
gauge invariant phase $\delta$ across the bridge. When $l/\xi(T)\ll1,$\ the
nanobridge behaves like a regular Josephson junction (JJ) with a sinusoidal
CPR\cite{Golubov_411}. However, as the ratio $l/\xi(T)$ becomes larger, the
CPR deviates from the sinosoidal form and can also become multivalued
\cite{Likharev_101}. In case the CPR is not multivalued the bridge can be
approximately considered as a JJ having an extra kinetic inductance
$L_{\mathrm{K}}$ . The effect of the kinetic inductance can be taken into
account by replacing the screening parameter of the loop $\beta_{L}%
=2\pi\Lambda I_{c}/\Phi_{0}$ by an effective one given by $\beta_{L}%
+\Delta\beta$, where $\Delta\beta=2\pi L_{\mathrm{K}}I_{\mathrm{c}}/\Phi_{0}$.

\bigskip In order to estimate $\Delta\beta$ we use Eqs. (47)-(49) and the data
in Fig. 5 of Ref. \cite{Troeman_024509}. For $l/\xi=1.7$ the bridges'
contribution is $\Delta\beta\simeq1$ . As we will discuss below, both
$\beta_{L}$ and $\Delta\beta$ depend on the injected power $P_{\mathrm{in}}$
that is used to excite the resonator due to a heating effect. However, for all
values of $P_{\mathrm{in}}$ that were used in our experiment, we estimate that
the ratio $\Delta\beta/\beta_{L}$ never exceeds the value $0.5$ and thus the
effect of kinetic inductance can be considered as small. Furthermore, the CPR
remains a single valued function in the entire range of parameters that is
explored in our experiments. Consequently, the nanobridges can be treated as
regular JJs to a good approximation.

\subsection{The SQUID Hamiltonian}

In the following derivation we treat the nanobridges as regular JJs. We
consider the case where the critical currents of both nanobridges are
$I_{\mathrm{c1}}=I_{\mathrm{c}}\left(  1+\alpha\right)  $ and $I_{\mathrm{c2}%
}=I_{\mathrm{c}}\left(  1-\alpha\right)  $ respectively, where the
dimensionless parameter $\alpha$ characterizes the asymmetry in the SQUID. The
Hamiltonian for the SQUID, which is expressed in terms of the two gauge
invariant phases $\delta_{1}$ and $\delta_{2}$ across both junctions, and
their canonical conjugates $p_{1}$ and $p_{2}$, is given by%
\begin{equation}
\mathcal{H}_{\mathrm{S}}\left(  I\right)  =\frac{2\pi\omega_{p}^{2}\left(
p_{1}^{2}+p_{2}^{2}\right)  }{E_{0}}+E_{0}u\left(  \delta_{1},\delta
_{2};I\right)  \;,
\end{equation}
where $\omega_{pl}=\sqrt{I_{c}/C_{\mathrm{J}}\Phi_{0}}$ is the plasma
frequency, $E_{0}=\Phi_{0}I_{c}/\pi$ is the Josephson energy, and the
dimensionless potential $u$ is given by \cite{Mitra_214512}%
\begin{equation}
u=-\frac{\left(  1+\alpha\right)  \cos\delta_{1}+\left(  1-\alpha\right)
\cos\delta_{2}}{2}+\frac{\left(  \frac{\delta_{1}-\delta_{2}}{2}+\frac{\pi
\Phi_{x}}{\Phi_{0}}\right)  ^{2}}{\beta_{L}}-\frac{\left(  \delta_{1}%
+\delta_{2}\right)  I}{4I_{c}}-\frac{\zeta\left(  \delta_{1}+\delta
_{2}\right)  ^{2}}{16}\;, \label{u=}%
\end{equation}
where $\zeta=\Phi_{0}/2I_{c}L_{\mathrm{T}}l_{\mathrm{T}}$.

\subsection{Adiabatic approximation}

Due to the extremely small capacitance $C_{\mathrm{J}}$ of both nanobridges
\cite{Ralph_10753}, the plasma frequency $\omega_{pl}$ of the SQUID is
estimated to exceed $1%
\operatorname{THz}%
$. Thus, the effect of the SQUID on the SSR, which has a much slower dynamics,
can be treated using the adiabatic approximation \cite{Buks_10001,Buks_026217}%
. Formally, treating the current $I$ as a parameter (rather than a degree of
freedom), the Hamiltonian $\mathcal{H}_{\mathrm{S}}$ can be diagonalized
$\mathcal{H}_{\mathrm{S}}\left\vert k\left(  I\right)  \right\rangle
=\varepsilon_{k}\left(  I\right)  \left\vert k\left(  I\right)  \right\rangle
$, where $k=0,1,2,...$, and $\left\langle k\left(  I\right)  |l\left(
I\right)  \right\rangle =\delta_{kl}$. To lowest order in the adiabatic
expansion the effective Hamiltonian governing the dynamics of the slow degrees
of freedom corresponding to the fast part of the system occupying the state
$\left\vert k\left(  I\right)  \right\rangle $ is given by $\mathcal{H}%
_{k}^{\mathrm{A}}=\mathcal{H}_{\mathrm{SSR}}+\varepsilon_{k}\left(  I\right)
$ \cite{Littlejohn_5239,Panati_250405}. Furthermore, in the limit where the
thermal energy $k_{\mathrm{B}}T$ is much smaller than the typical energy
spacing between different levels of $\mathcal{H}_{1}$ ($\simeq\hbar\omega
_{pl}$) one can assume that the SQUID remains in its current dependent ground
state $\left\vert 0\left(  I\right)  \right\rangle $. For most cases this
assumption is valid for our experimental parameters. It is important, however,
to note that when the externally applied magnetic flux is close to a
half-integer value (in units of $\Phi_{0}$), namely, when $\Phi_{x}%
\simeq\left(  n+1/2\right)  \Phi_{0}$, where $n$ is integer, this
approximation may break down. Near these points the potential $u$ may have two
different neighboring wells having similar depth. Consequently, near these
points, the energy gap between the ground state and the first excited state
can become much smaller than $\hbar\omega_{pl}$. On the other hand, the ratio
between the height of the barrier separating the two wells ($\simeq E_{0}$)
and the energy spacing between intra-well states ($\simeq\hbar\omega_{pl}$) is
typically $E_{0}/\hbar\omega_{pl}\simeq100$ for our samples. Since the
coupling between states localized in different wells depends exponentially on
this ratio, we conclude that to a good approximation the inter-well coupling
can be neglected. Moreover, in the same limit where $E_{0}/\hbar\omega_{pl}%
\gg1$, one can approximate the ground state energy $\varepsilon_{0}$ by the
value of $E_{0}u$ at the bottom of the well where the system is localized.

The current $I$ at the center of the SSR can readably be expressed in terms of
the annihilation and creation operators $A_{1}$, $A_{1}^{\dag}$ $A_{3}$ and
$A_{3}^{\dag}$. This allows expanding the current dependent ground state
energy $\varepsilon_{0}\left(  I\right)  $ as a power series of these
operators. In the rotating wave approximation oscillating terms in such an
expansion are neglected since their effect on the dynamics for a time scale
much longer than a typical oscillation period is negligibly small. Moreover,
constant terms in the Hamiltonian are disregarded since they only give rise to
a global phase factor. In the present experiment the 1st SSR mode is
externally driven, and we focus on the resultant dephasing induced on the 3rd
mode. To that end we include in the effective Hamiltonian of the closed system
in addition to the linear terms corresponding to the 1st and 3rd modes, also
the Kerr nonlinearity term of the 1st mode, which is externally driven, and
also the term representing intermode coupling between the 1st and the 3rd
modes%
\begin{equation}
\mathcal{H}_{\mathrm{eff}}=\hbar\omega_{1}N_{1}+\hbar\omega_{3}N_{3}%
+V_{\mathrm{in}}+\hbar K_{1}N_{1}^{2}+\hbar\lambda_{1,3}N_{1}N_{3}\ .
\label{H_eff}%
\end{equation}
The angular resonance frequency shift of the 1st and the 3rd modes, which is
given by%
\begin{equation}
\frac{\omega_{1}-\omega_{\mathrm{T}}}{\omega_{\mathrm{T}}}=\frac{\omega
_{3}-3\omega_{\mathrm{T}}}{3\omega_{\mathrm{T}}}=\zeta\frac{\partial
^{2}\left(  \varepsilon_{0}/E_{0}\right)  }{\partial\left(  I/I_{\mathrm{c}%
}\right)  ^{2}}\;,
\end{equation}
can be attributed to the inductance of the SQUID, which is proportional to the
second derivative of $\varepsilon_{0}$ with respect to $I$. On the other hand,
the Kerr nonlinearity, which is given by%
\begin{equation}
\frac{K_{1}}{\omega_{1}}=\frac{\zeta^{2}\hbar\omega_{1}}{2E_{0}}\frac
{\partial^{4}\left(  \varepsilon_{0}/E_{0}\right)  }{\partial\left(
I/I_{\mathrm{c}}\right)  ^{4}}\;,
\end{equation}
and the intermode coupling, which is given by $\lambda_{1,3}=9K_{1}$, can both
be attributed to the nonlinear inductance of the SQUID \cite{Yurke_5054},
which is proportional to the fourth derivative of $\varepsilon_{0}$ with
respect to $I$.

\subsection{Evaluation of $\omega_{1},$ $\omega_{3},$ $K_{1}$ and
$\lambda_{1,3}$ in the limit $\beta_{L}\ll1$}

The evaluation of the parameters $\omega_{1}$, $\omega_{3}$, $K_{1}$ and
$\lambda_{1,3}$ generally requires a numerical calculation. However, an
analytical approximation can be employed when $\beta_{L}\ll1$. In this limit
the phase difference $\delta_{2}-\delta_{1}$ is strongly confined near the
value $2\pi\Phi_{x}/\Phi_{0}$, as can be seen from Eq. (\ref{u=}). This fact
can be exploited to further simplify the dynamics by applying another
adiabatic approximation, in which the phase difference $\delta_{2}-\delta_{1}$
is treated as a 'fast' variable and the phase average $\delta_{+}=\left(
\delta_{1}+\delta_{2}\right)  /2$ as a 'slow' one. To lowest order in the
adiabatic expansion one finds that for low frequencies $\omega\ll\omega_{pl}$,
namely in the region where the impedance associated with the capacitance of
the JJs is much larger in absolute value in comparison with the impedance
associated with the inductance, the SQUID behaves as a single JJ having
critical current given by \cite{Tesche_380}%
\begin{equation}
I_{\mathrm{S}}=2I_{\mathrm{c}}\sqrt{1-\left(  1-\alpha^{2}\right)  \sin
^{2}\left(  \pi\Phi_{x}/\Phi_{0}\right)  }\;.
\end{equation}
Note that this approximation may break down when $\Phi_{x}\simeq\left(
n+1/2\right)  \Phi_{0}$ unless the asymmetry parameter $\alpha$ is
sufficiently large. The relatively large value of $\alpha$ in our device
($\alpha\simeq0.5$) ensures the validity of the above approximation. Using
this result, it is straightforward to obtain the following analytical approximations:%

\begin{subequations}
\begin{align}
\frac{\partial^{2}\left(  \varepsilon_{0}/E_{0}\right)  }{\partial\left(
I/I_{\mathrm{c}}\right)  ^{2}}  &  =\frac{I_{\mathrm{c}}}{\pi I_{\mathrm{S}}%
}\;,\\
\frac{\partial^{4}\left(  \varepsilon_{0}/E_{0}\right)  }{\partial\left(
I/I_{\mathrm{c}}\right)  ^{4}}  &  =-\frac{8}{3\pi^{2}}\left(  \frac
{I_{\mathrm{c}}}{I_{\mathrm{S}}}\right)  ^{3}\;,
\end{align}
which can be used to evaluate all the terms in Eq. (\ref{H_eff}).

\section{Hysteretic response and heating of the nanobridges}

As we discuss in \cite{Suchoi_3133}, the resonator exhibits hysteretic
response to magnetic flux when the input power is relatively low. Such a
behavior occurs, as can be seen from Eq. (\ref{u=}) above, when the screening
parameter $\beta_{L}$ is sufficiently large to give rise to metastability in
the dimensionless potential $u$. A fitting of the model to the experimental
data shown in Fig. 3(a) of \cite{Suchoi_3133} yields a value of $\beta
_{L}=7.4$. Another example of hysteretic response is shown in Fig.
\ref{largebetal} below that shows data taken with another sample, which was
fabricated using the same process that is described in the first section. The
larger critical current in that sample yields a larger value of the screening
parameter $\beta_{L}=20$.%

\begin{figure}
[b]
\begin{center}
\includegraphics[
height=2.0003in,
width=3in
]%
{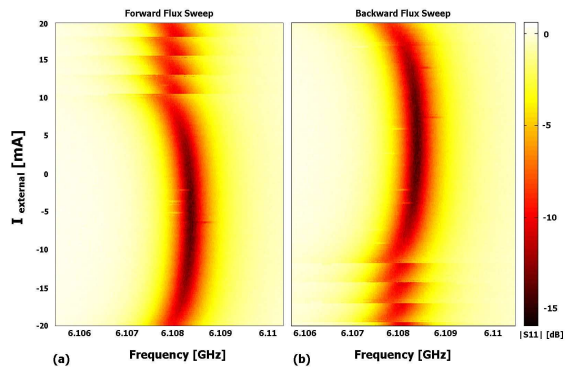}%
\caption{Measured $\left\vert S_{11}\right\vert $ at input power
$P_{\mathrm{in}}=-95$dBm for forward (a) and backward (b) magnetic flux sweep.
In this sample $\beta_{L}=20,$ and the response is highly hysteretic.}%
\label{largebetal}%
\end{center}
\end{figure}

As is mentioned in the \cite{Suchoi_3133}, as the input power is increased the
response becomes non-hysteretic. The gradual transition between the hysteretic
region to the non-hysteretic one is seen in Fig. \ref{inc_dec_dif} below,
which shows the difference in the measured resonance frequency of the first
mode obtained from increased flux sweep $\left(  f_{\mathrm{1inc}}\right)  $
and decreased flux sweep $\left(  f_{\mathrm{1dec}}\right)  $ at different
input powers. Dark blue in the color map corresponds to no difference, namely
to monostable regions, whereas in the red regions, where a large difference is
observed, the system is bistable. As can be clearly seen from the figure, the
bistable regions shrink as the input power is increased. The experimental
results suggest that the critical current of the nanobridges drops as the
input power is increased, and consequently the response becomes non-hysteretic
due to the resultant smaller value of the screening parameter $\beta_{L}$. We
hypothesize that the drop in the critical current occurs due to heating of the
nanobridges by the input power.%

\begin{figure}
[b]
\begin{center}
\includegraphics[
trim=0.000000in 0.000000in -0.013605in 0.000000in,
height=2.4241in,
width=3.2335in
]%
{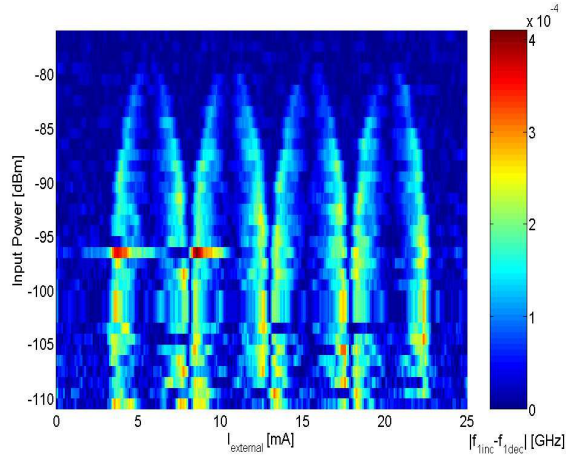}%
\caption{The difference between the measured resonance frequencies obtained in
the increasing flux sweep ($f_{\mathrm{1inc}})$ and the decreasing flux sweep
($f_{\mathrm{1dec}})$ of the first (\textit{detector}) mode. The dark blue
areas correspond to monostable regions, namely, the same resonance frequency
is measured for both the increased and decreased sweep. The red indicates the
regions where the system is bistable.}%
\label{inc_dec_dif}%
\end{center}
\end{figure}

To estimate the effect of heating, we assume the case where the substrate is
isothermal and that the heat is mainly dissipated down into the substrate
rather than along the film \cite{Johnson_7069}. Moreover, we assume that most
of the externally injected power into the resonator is dissipated near the
nanobridges, where, the current density obtains its largest value. By
estimating the heat transfer coefficient per unit area between each nanobridge
and the substrate beneath it ($100%
\operatorname{nm}%
$ SiN on top of high-resistivity Si) to be $\kappa\simeq1%
\operatorname{W}%
\operatorname{cm}%
^{-2}%
\operatorname{K}%
^{-1}$ \cite{Monticone_3866,Weiser_4888} and the area of the nanobridge to be
$A\simeq\left(  50%
\operatorname{nm}%
\right)  ^{2}$one finds that the expected temperature rise for $P_{\mathrm{in}%
}=-70$ dBm is $\Delta T=P_{\mathrm{in}}/A\kappa\simeq4%
\operatorname{K}%
$.

Since heating is produced by AC current flowing through the nanobridges, it is
important to estimate also the thermal rate, which characterizes the inverse
of the typical time scale of thermalization, and is given by $\gamma
_{\mathrm{T}}=A\kappa/C$, where the heat capacity $C$ of the nanobridge is
given by $C=C_{v}Ad$, $C_{v}$ is the heat capacity per unit volume, and $d$ is
the thickness of the superconducting film. Using the estimate $C_{v}%
\simeq10^{-3}%
\operatorname{J}%
\operatorname{cm}%
^{-3}%
\operatorname{K}%
^{-1}$ \cite{Weiser_4888} one finds $\gamma_{\mathrm{T}}\simeq0.1%
\operatorname{GHz}%
$. Since the frequency of the AC heating current is 1-2 orders of magnitude
higher, we conclude that to a good approximation the temperature of the
nanobridges can be considered as stationary in the steady state.\newpage
\bibliographystyle{ieee}
\bibliography{acompat,Eyal_Bib}

\newif\ifabfull\abfulltrue
\begin{thebibliography}{33}
\expandafter\ifx\csname natexlab\endcsname\relax\def\natexlab#1{#1}\fi
\expandafter\ifx\csname bibnamefont\endcsname\relax
  \def\bibnamefont#1{#1}\fi
\expandafter\ifx\csname bibfnamefont\endcsname\relax
  \def\bibfnamefont#1{#1}\fi
\expandafter\ifx\csname citenamefont\endcsname\relax
  \def\citenamefont#1{#1}\fi
\expandafter\ifx\csname url\endcsname\relax
  \def\url#1{\texttt{#1}}\fi
\expandafter\ifx\csname urlprefix\endcsname\relax\def\urlprefix{URL }\fi
\providecommand{\bibinfo}[2]{#2}
\providecommand{\eprint}[2][]{\url{#2}}

\bibitem[{\citenamefont{Suchoi et~al.}(2009)\citenamefont{Suchoi, Abdo, Segev,
  Shtempluck, Blencowe, and Buks}}]{Suchoi_3133}
\bibinfo{author}{\bibfnamefont{O.}~\bibnamefont{Suchoi}},
  \bibinfo{author}{\bibfnamefont{B.}~\bibnamefont{Abdo}},
  \bibinfo{author}{\bibfnamefont{E.}~\bibnamefont{Segev}},
  \bibinfo{author}{\bibfnamefont{O.}~\bibnamefont{Shtempluck}},
  \bibinfo{author}{\bibfnamefont{M.~P.} \bibnamefont{Blencowe}},
  \bibnamefont{and} \bibinfo{author}{\bibfnamefont{E.}~\bibnamefont{Buks}},
  \bibinfo{journal}{arXiv:0901.3133v1}  (\bibinfo{year}{2009}).

\bibitem[{\citenamefont{Hao et~al.}(2008)\citenamefont{Hao, Macfarlane, Gallop,
  Cox, Beyer, Drung, and Schurig}}]{Hao_192507}
\bibinfo{author}{\bibfnamefont{L.}~\bibnamefont{Hao}},
  \bibinfo{author}{\bibfnamefont{J.~C.} \bibnamefont{Macfarlane}},
  \bibinfo{author}{\bibfnamefont{J.~C.} \bibnamefont{Gallop}},
  \bibinfo{author}{\bibfnamefont{D.}~\bibnamefont{Cox}},
  \bibinfo{author}{\bibfnamefont{J.}~\bibnamefont{Beyer}},
  \bibinfo{author}{\bibfnamefont{D.}~\bibnamefont{Drung}}, \bibnamefont{and}
  \bibinfo{author}{\bibfnamefont{T.}~\bibnamefont{Schurig}},
  \bibinfo{journal}{Applied Physics Letters} \textbf{\bibinfo{volume}{92}},
  \bibinfo{eid}{192507} (pages~\bibinfo{numpages}{3}) (\bibinfo{year}{2008}),
  \urlprefix\url{http://link.aip.org/link/?APL/92/192507/1}.

\bibitem[{\citenamefont{Hao et~al.}(2007)\citenamefont{Hao, Macfarlane, Gallop,
  Cox, Joseph-Franks, Hutson, Chen, and Lam}}]{Hao_392}
\bibinfo{author}{\bibfnamefont{L.}~\bibnamefont{Hao}},
  \bibinfo{author}{\bibfnamefont{J.~C.} \bibnamefont{Macfarlane}},
  \bibinfo{author}{\bibfnamefont{J.~C.} \bibnamefont{Gallop}},
  \bibinfo{author}{\bibfnamefont{D.}~\bibnamefont{Cox}},
  \bibinfo{author}{\bibfnamefont{P.}~\bibnamefont{Joseph-Franks}},
  \bibinfo{author}{\bibfnamefont{D.}~\bibnamefont{Hutson}},
  \bibinfo{author}{\bibfnamefont{J.}~\bibnamefont{Chen}}, \bibnamefont{and}
  \bibinfo{author}{\bibfnamefont{S.~K.~H.} \bibnamefont{Lam}},
  \bibinfo{journal}{IEEE Transactions on Instrumentation and Measurement}
  \textbf{\bibinfo{volume}{56}}, \bibinfo{pages}{392} (\bibinfo{year}{2007}).

\bibitem[{\citenamefont{Bell et~al.}(2003)\citenamefont{Bell, Burnell, Kang,
  Hadfield, Kappers, and Blamire}}]{Bell_630}
\bibinfo{author}{\bibfnamefont{C.}~\bibnamefont{Bell}},
  \bibinfo{author}{\bibfnamefont{G.}~\bibnamefont{Burnell}},
  \bibinfo{author}{\bibfnamefont{D.-J.} \bibnamefont{Kang}},
  \bibinfo{author}{\bibfnamefont{R.~H.} \bibnamefont{Hadfield}},
  \bibinfo{author}{\bibfnamefont{M.~J.} \bibnamefont{Kappers}},
  \bibnamefont{and} \bibinfo{author}{\bibfnamefont{M.~G.}
  \bibnamefont{Blamire}}, \bibinfo{journal}{Nanotechnology}
  \textbf{\bibinfo{volume}{14}}, \bibinfo{pages}{630} (\bibinfo{year}{2003}),
  \urlprefix\url{http://stacks.iop.org/0957-4484/14/630}.

\bibitem[{\citenamefont{Datesman
  et~al.}(2005{\natexlab{a}})\citenamefont{Datesman, Schultz, Lichtenberger,
  Golish, Walker, and Kooi}}]{Datesman_928}
\bibinfo{author}{\bibfnamefont{A.}~\bibnamefont{Datesman}},
  \bibinfo{author}{\bibfnamefont{J.}~\bibnamefont{Schultz}},
  \bibinfo{author}{\bibfnamefont{A.}~\bibnamefont{Lichtenberger}},
  \bibinfo{author}{\bibfnamefont{D.}~\bibnamefont{Golish}},
  \bibinfo{author}{\bibfnamefont{C.}~\bibnamefont{Walker}}, \bibnamefont{and}
  \bibinfo{author}{\bibfnamefont{J.}~\bibnamefont{Kooi}},
  \bibinfo{journal}{IEEE Transactions on Applied Superconductivity}
  \textbf{\bibinfo{volume}{15}}, \bibinfo{pages}{928}
  (\bibinfo{year}{2005}{\natexlab{a}}).

\bibitem[{\citenamefont{Troeman et~al.}(2007)\citenamefont{Troeman, Derking,
  Borger, Pleikies, Veldhuis, and Hilgenkamp}}]{Troeman_2152}
\bibinfo{author}{\bibfnamefont{A.}~\bibnamefont{Troeman}},
  \bibinfo{author}{\bibfnamefont{H.}~\bibnamefont{Derking}},
  \bibinfo{author}{\bibfnamefont{B.}~\bibnamefont{Borger}},
  \bibinfo{author}{\bibfnamefont{J.}~\bibnamefont{Pleikies}},
  \bibinfo{author}{\bibfnamefont{D.}~\bibnamefont{Veldhuis}}, \bibnamefont{and}
  \bibinfo{author}{\bibfnamefont{H.}~\bibnamefont{Hilgenkamp}},
  \bibinfo{journal}{Nano Letters} \textbf{\bibinfo{volume}{7}},
  \bibinfo{pages}{2152} (\bibinfo{year}{2007}), ISSN \bibinfo{issn}{1530-6984},
  \urlprefix\url{http://pubs3.acs.org/acs/journals/doilookup?in_doi=10.1021/nl%
070870f}.

\bibitem[{\citenamefont{Datesman
  et~al.}(2005{\natexlab{b}})\citenamefont{Datesman, Schultz, Cecil, Lyons, and
  Lichtenberger}}]{Datesman_3524}
\bibinfo{author}{\bibfnamefont{A.}~\bibnamefont{Datesman}},
  \bibinfo{author}{\bibfnamefont{J.}~\bibnamefont{Schultz}},
  \bibinfo{author}{\bibfnamefont{T.}~\bibnamefont{Cecil}},
  \bibinfo{author}{\bibfnamefont{C.}~\bibnamefont{Lyons}}, \bibnamefont{and}
  \bibinfo{author}{\bibfnamefont{A.}~\bibnamefont{Lichtenberger}},
  \bibinfo{journal}{IEEE Transactions on Applied Superconductivity}
  \textbf{\bibinfo{volume}{15}}, \bibinfo{pages}{3524}
  (\bibinfo{year}{2005}{\natexlab{b}}), ISSN \bibinfo{issn}{1051-8223}.

\bibitem[{\citenamefont{Clark et~al.}(2001)\citenamefont{Clark, Prance,
  Whiteman, Prance, Everitt, Bulsara, and Ralph}}]{Clark_3042}
\bibinfo{author}{\bibfnamefont{T.~D.} \bibnamefont{Clark}},
  \bibinfo{author}{\bibfnamefont{R.~J.} \bibnamefont{Prance}},
  \bibinfo{author}{\bibfnamefont{R.}~\bibnamefont{Whiteman}},
  \bibinfo{author}{\bibfnamefont{H.}~\bibnamefont{Prance}},
  \bibinfo{author}{\bibfnamefont{M.~J.} \bibnamefont{Everitt}},
  \bibinfo{author}{\bibfnamefont{A.~R.} \bibnamefont{Bulsara}},
  \bibnamefont{and} \bibinfo{author}{\bibfnamefont{J.~F.} \bibnamefont{Ralph}},
  \bibinfo{journal}{Journal of Applied Physics} \textbf{\bibinfo{volume}{90}},
  \bibinfo{pages}{3042} (\bibinfo{year}{2001}),
  \urlprefix\url{http://link.aip.org/link/?JAP/90/3042/1}.

\bibitem[{\citenamefont{Blencowe and Buks}(2007)}]{Blencowe_014511}
\bibinfo{author}{\bibfnamefont{M.~P.} \bibnamefont{Blencowe}} \bibnamefont{and}
  \bibinfo{author}{\bibfnamefont{E.}~\bibnamefont{Buks}},
  \bibinfo{journal}{Phys. Rev. B} \textbf{\bibinfo{volume}{76}},
  \bibinfo{pages}{14511} (\bibinfo{year}{2007}).

\bibitem[{\citenamefont{Nation et~al.}(2008)\citenamefont{Nation, Blencowe, and
  Buks}}]{Nation_104516}
\bibinfo{author}{\bibfnamefont{P.~D.} \bibnamefont{Nation}},
  \bibinfo{author}{\bibfnamefont{M.~P.} \bibnamefont{Blencowe}},
  \bibnamefont{and} \bibinfo{author}{\bibfnamefont{E.}~\bibnamefont{Buks}},
  \bibinfo{journal}{Phys. Rev. B} \textbf{\bibinfo{volume}{78}},
  \bibinfo{pages}{104516} (\bibinfo{year}{2008}).

\bibitem[{\citenamefont{Granata et~al.}(2008)\citenamefont{Granata, Esposito,
  Vettoliere, Petti, and Russo}}]{Granata_275501}
\bibinfo{author}{\bibfnamefont{C.}~\bibnamefont{Granata}},
  \bibinfo{author}{\bibfnamefont{E.}~\bibnamefont{Esposito}},
  \bibinfo{author}{\bibfnamefont{A.}~\bibnamefont{Vettoliere}},
  \bibinfo{author}{\bibfnamefont{L.}~\bibnamefont{Petti}}, \bibnamefont{and}
  \bibinfo{author}{\bibfnamefont{M.}~\bibnamefont{Russo}},
  \bibinfo{journal}{Nanotechnology} \textbf{\bibinfo{volume}{19}},
  \bibinfo{pages}{275501} (\bibinfo{year}{2008}).

\bibitem[{\citenamefont{Hasselbach et~al.}(2002)\citenamefont{Hasselbach,
  Mailly, and Kirtley}}]{Hasselbach_4432}
\bibinfo{author}{\bibfnamefont{K.}~\bibnamefont{Hasselbach}},
  \bibinfo{author}{\bibfnamefont{D.}~\bibnamefont{Mailly}}, \bibnamefont{and}
  \bibinfo{author}{\bibfnamefont{J.}~\bibnamefont{Kirtley}},
  \bibinfo{journal}{Journal of Applied Physics} \textbf{\bibinfo{volume}{91}},
  \bibinfo{pages}{4432} (\bibinfo{year}{2002}).

\bibitem[{\citenamefont{Hasselbach et~al.}(2000)\citenamefont{Hasselbach,
  Veauvy, and Mailly}}]{Hasselbach_140}
\bibinfo{author}{\bibfnamefont{K.}~\bibnamefont{Hasselbach}},
  \bibinfo{author}{\bibfnamefont{C.}~\bibnamefont{Veauvy}}, \bibnamefont{and}
  \bibinfo{author}{\bibfnamefont{D.}~\bibnamefont{Mailly}},
  \bibinfo{journal}{Physica C Superconductivity}
  \textbf{\bibinfo{volume}{332}}, \bibinfo{pages}{140} (\bibinfo{year}{2000}).

\bibitem[{\citenamefont{Baratoff et~al.}(1970)\citenamefont{Baratoff,
  Blackburn, and Schwartz}}]{Baratoff_1096}
\bibinfo{author}{\bibfnamefont{A.}~\bibnamefont{Baratoff}},
  \bibinfo{author}{\bibfnamefont{J.~A.} \bibnamefont{Blackburn}},
  \bibnamefont{and} \bibinfo{author}{\bibfnamefont{B.~B.}
  \bibnamefont{Schwartz}}, \bibinfo{journal}{Phys. Rev. Lett.}
  \textbf{\bibinfo{volume}{25}}, \bibinfo{pages}{1096} (\bibinfo{year}{1970}).

\bibitem[{\citenamefont{Likharev}(1979)}]{Likharev_101}
\bibinfo{author}{\bibfnamefont{K.~K.} \bibnamefont{Likharev}},
  \bibinfo{journal}{Rev. Mod. Phys.} \textbf{\bibinfo{volume}{51}},
  \bibinfo{pages}{101} (\bibinfo{year}{1979}).

\bibitem[{\citenamefont{Likharev and Yakobson}(1975)}]{Likharev_950}
\bibinfo{author}{\bibfnamefont{K.~K.} \bibnamefont{Likharev}} \bibnamefont{and}
  \bibinfo{author}{\bibfnamefont{L.~A.} \bibnamefont{Yakobson}},
  \bibinfo{journal}{Sov. Phys. - Tech. Phys. (Engl. Transl.)}
  \textbf{\bibinfo{volume}{20}}, \bibinfo{pages}{950} (\bibinfo{year}{1975}).

\bibitem[{\citenamefont{Gumann et~al.}(2007)\citenamefont{Gumann, Dahm, and
  Schopohl}}]{Gumann_064529}
\bibinfo{author}{\bibfnamefont{A.}~\bibnamefont{Gumann}},
  \bibinfo{author}{\bibfnamefont{T.}~\bibnamefont{Dahm}}, \bibnamefont{and}
  \bibinfo{author}{\bibfnamefont{N.}~\bibnamefont{Schopohl}},
  \bibinfo{journal}{Physical Review B (Condensed Matter and Materials Physics)}
  \textbf{\bibinfo{volume}{76}}, \bibinfo{eid}{064529}
  (pages~\bibinfo{numpages}{14}) (\bibinfo{year}{2007}),
  \urlprefix\url{http://link.aps.org/abstract/PRB/v76/e064529}.

\bibitem[{\citenamefont{Podd et~al.}(2007)\citenamefont{Podd, Hutchinson,
  Williams, and Hasko}}]{Podd_134501}
\bibinfo{author}{\bibfnamefont{G.~J.} \bibnamefont{Podd}},
  \bibinfo{author}{\bibfnamefont{G.~D.} \bibnamefont{Hutchinson}},
  \bibinfo{author}{\bibfnamefont{D.~A.} \bibnamefont{Williams}},
  \bibnamefont{and} \bibinfo{author}{\bibfnamefont{D.~G.} \bibnamefont{Hasko}},
  \bibinfo{journal}{Physical Review B (Condensed Matter and Materials Physics)}
  \textbf{\bibinfo{volume}{75}}, \bibinfo{pages}{134501}
  (\bibinfo{year}{2007}).

\bibitem[{\citenamefont{Pronin et~al.}(1998)\citenamefont{Pronin, Dressel,
  Pimenov, Loidl, Roshchin, and Greene}}]{Pronin_14416}
\bibinfo{author}{\bibfnamefont{A.~V.} \bibnamefont{Pronin}},
  \bibinfo{author}{\bibfnamefont{M.}~\bibnamefont{Dressel}},
  \bibinfo{author}{\bibfnamefont{A.}~\bibnamefont{Pimenov}},
  \bibinfo{author}{\bibfnamefont{A.}~\bibnamefont{Loidl}},
  \bibinfo{author}{\bibfnamefont{I.~V.} \bibnamefont{Roshchin}},
  \bibnamefont{and} \bibinfo{author}{\bibfnamefont{L.~H.}
  \bibnamefont{Greene}}, \bibinfo{journal}{Phys. Rev. B}
  \textbf{\bibinfo{volume}{57}}, \bibinfo{pages}{14416} (\bibinfo{year}{1998}).

\bibitem[{\citenamefont{Maxfield and McLean}(1965)}]{Maxfield_A1515}
\bibinfo{author}{\bibfnamefont{B.~W.} \bibnamefont{Maxfield}} \bibnamefont{and}
  \bibinfo{author}{\bibfnamefont{W.~L.} \bibnamefont{McLean}},
  \bibinfo{journal}{Phys. Rev.} \textbf{\bibinfo{volume}{139}},
  \bibinfo{pages}{A1515} (\bibinfo{year}{1965}).

\bibitem[{\citenamefont{Golubov et~al.}(2004)\citenamefont{Golubov, Kupriyanov,
  and Il\char39{}ichev}}]{Golubov_411}
\bibinfo{author}{\bibfnamefont{A.~A.} \bibnamefont{Golubov}},
  \bibinfo{author}{\bibfnamefont{M.~Y.} \bibnamefont{Kupriyanov}},
  \bibnamefont{and}
  \bibinfo{author}{\bibfnamefont{E.}~\bibnamefont{Il\char39{}ichev}},
  \bibinfo{journal}{Rev. Mod. Phys.} \textbf{\bibinfo{volume}{76}},
  \bibinfo{pages}{411} (\bibinfo{year}{2004}).

\bibitem[{\citenamefont{Troeman et~al.}(2008)\citenamefont{Troeman, van~der
  Ploeg, Il'Ichev, Meyer, Golubov, Kupriyanov, and
  Hilgenkamp}}]{Troeman_024509}
\bibinfo{author}{\bibfnamefont{A.~G.~P.} \bibnamefont{Troeman}},
  \bibinfo{author}{\bibfnamefont{S.~H.~W.} \bibnamefont{van~der Ploeg}},
  \bibinfo{author}{\bibfnamefont{E.}~\bibnamefont{Il'Ichev}},
  \bibinfo{author}{\bibfnamefont{H.-G.} \bibnamefont{Meyer}},
  \bibinfo{author}{\bibfnamefont{A.~A.} \bibnamefont{Golubov}},
  \bibinfo{author}{\bibfnamefont{M.~Y.} \bibnamefont{Kupriyanov}},
  \bibnamefont{and}
  \bibinfo{author}{\bibfnamefont{H.}~\bibnamefont{Hilgenkamp}},
  \bibinfo{journal}{Physical Review B (Condensed Matter and Materials Physics)}
  \textbf{\bibinfo{volume}{77}}, \bibinfo{eid}{024509}
  (pages~\bibinfo{numpages}{5}) (\bibinfo{year}{2008}),
  \urlprefix\url{http://link.aps.org/abstract/PRB/v77/e024509}.

\bibitem[{\citenamefont{Mitra et~al.}(2008)\citenamefont{Mitra, Strauch, Lobb,
  Anderson, Wellstood, and Tiesinga}}]{Mitra_214512}
\bibinfo{author}{\bibfnamefont{K.}~\bibnamefont{Mitra}},
  \bibinfo{author}{\bibfnamefont{F.~W.} \bibnamefont{Strauch}},
  \bibinfo{author}{\bibfnamefont{C.~J.} \bibnamefont{Lobb}},
  \bibinfo{author}{\bibfnamefont{J.~R.} \bibnamefont{Anderson}},
  \bibinfo{author}{\bibfnamefont{F.~C.} \bibnamefont{Wellstood}},
  \bibnamefont{and} \bibinfo{author}{\bibfnamefont{E.}~\bibnamefont{Tiesinga}},
  \bibinfo{journal}{Physical Review B (Condensed Matter and Materials Physics)}
  \textbf{\bibinfo{volume}{77}}, \bibinfo{eid}{214512}
  (pages~\bibinfo{numpages}{10}) (\bibinfo{year}{2008}),
  \urlprefix\url{http://link.aps.org/abstract/PRB/v77/e214512}.

\bibitem[{\citenamefont{Ralph et~al.}(1996)\citenamefont{Ralph, Clark, Prance,
  Prance, and Diggins}}]{Ralph_10753}
\bibinfo{author}{\bibfnamefont{J.~F.} \bibnamefont{Ralph}},
  \bibinfo{author}{\bibfnamefont{T.~D.} \bibnamefont{Clark}},
  \bibinfo{author}{\bibfnamefont{R.~J.} \bibnamefont{Prance}},
  \bibinfo{author}{\bibfnamefont{H.}~\bibnamefont{Prance}}, \bibnamefont{and}
  \bibinfo{author}{\bibfnamefont{J.}~\bibnamefont{Diggins}},
  \bibinfo{journal}{J. Phys.: Condens. Matter} \textbf{\bibinfo{volume}{8}},
  \bibinfo{pages}{10753} (\bibinfo{year}{1996}).

\bibitem[{\citenamefont{Buks et~al.}(2008)\citenamefont{Buks, Arbel-Segev,
  Zaitsev, Abdo, and Blencowe}}]{Buks_10001}
\bibinfo{author}{\bibfnamefont{E.}~\bibnamefont{Buks}},
  \bibinfo{author}{\bibfnamefont{E.}~\bibnamefont{Arbel-Segev}},
  \bibinfo{author}{\bibfnamefont{S.}~\bibnamefont{Zaitsev}},
  \bibinfo{author}{\bibfnamefont{B.}~\bibnamefont{Abdo}}, \bibnamefont{and}
  \bibinfo{author}{\bibfnamefont{M.~P.} \bibnamefont{Blencowe}},
  \bibinfo{journal}{Europhys. Lett.} \textbf{\bibinfo{volume}{81}},
  \bibinfo{pages}{10001} (\bibinfo{year}{2008}).

\bibitem[{\citenamefont{Buks et~al.}(2007)\citenamefont{Buks, Zaitsev, Segev,
  Abdo, and Blencowe}}]{Buks_026217}
\bibinfo{author}{\bibfnamefont{E.}~\bibnamefont{Buks}},
  \bibinfo{author}{\bibfnamefont{S.}~\bibnamefont{Zaitsev}},
  \bibinfo{author}{\bibfnamefont{E.}~\bibnamefont{Segev}},
  \bibinfo{author}{\bibfnamefont{B.}~\bibnamefont{Abdo}}, \bibnamefont{and}
  \bibinfo{author}{\bibfnamefont{M.~P.} \bibnamefont{Blencowe}},
  \bibinfo{journal}{Phys. Rev. E} \textbf{\bibinfo{volume}{76}},
  \bibinfo{pages}{26217} (\bibinfo{year}{2007}).

\bibitem[{\citenamefont{Littlejohn and Flynn}(1991)}]{Littlejohn_5239}
\bibinfo{author}{\bibfnamefont{R.~G.} \bibnamefont{Littlejohn}}
  \bibnamefont{and} \bibinfo{author}{\bibfnamefont{W.~G.} \bibnamefont{Flynn}},
  \bibinfo{journal}{Phys. Rev. A} \textbf{\bibinfo{volume}{44}},
  \bibinfo{pages}{5239} (\bibinfo{year}{1991}).

\bibitem[{\citenamefont{Panati et~al.}(2002)\citenamefont{Panati, Spohn, and
  Teufel}}]{Panati_250405}
\bibinfo{author}{\bibfnamefont{G.}~\bibnamefont{Panati}},
  \bibinfo{author}{\bibfnamefont{H.}~\bibnamefont{Spohn}}, \bibnamefont{and}
  \bibinfo{author}{\bibfnamefont{S.}~\bibnamefont{Teufel}},
  \bibinfo{journal}{Phys. Rev. Lett.} \textbf{\bibinfo{volume}{88}},
  \bibinfo{pages}{250405} (\bibinfo{year}{2002}).

\bibitem[{\citenamefont{Yurke and Buks}(2006)}]{Yurke_5054}
\bibinfo{author}{\bibfnamefont{B.}~\bibnamefont{Yurke}} \bibnamefont{and}
  \bibinfo{author}{\bibfnamefont{E.}~\bibnamefont{Buks}}, \bibinfo{journal}{J.
  Lightwave Tech.} \textbf{\bibinfo{volume}{24}}, \bibinfo{pages}{5054}
  (\bibinfo{year}{2006}).

\bibitem[{\citenamefont{Tesche and Clarke}(1977)}]{Tesche_380}
\bibinfo{author}{\bibfnamefont{C.~D.} \bibnamefont{Tesche}} \bibnamefont{and}
  \bibinfo{author}{\bibfnamefont{J.}~\bibnamefont{Clarke}},
  \bibinfo{journal}{J. low Temp. Phys.} \textbf{\bibinfo{volume}{29}},
  \bibinfo{pages}{301} (\bibinfo{year}{1977}).

\bibitem[{\citenamefont{Johnson et~al.}(1996)\citenamefont{Johnson, Herr, and
  Kadin}}]{Johnson_7069}
\bibinfo{author}{\bibfnamefont{M.~W.} \bibnamefont{Johnson}},
  \bibinfo{author}{\bibfnamefont{A.~M.} \bibnamefont{Herr}}, \bibnamefont{and}
  \bibinfo{author}{\bibfnamefont{A.~M.} \bibnamefont{Kadin}},
  \bibinfo{journal}{J. Appl. Phys.} \textbf{\bibinfo{volume}{79}},
  \bibinfo{pages}{7069} (\bibinfo{year}{1996}).

\bibitem[{\citenamefont{Monticone et~al.}(1999)\citenamefont{Monticone,
  Lacquaniti, Steni, Rajteri, Rastello, and Parlato}}]{Monticone_3866}
\bibinfo{author}{\bibfnamefont{E.}~\bibnamefont{Monticone}},
  \bibinfo{author}{\bibfnamefont{V.}~\bibnamefont{Lacquaniti}},
  \bibinfo{author}{\bibfnamefont{R.}~\bibnamefont{Steni}},
  \bibinfo{author}{\bibfnamefont{M.}~\bibnamefont{Rajteri}},
  \bibinfo{author}{\bibfnamefont{M.}~\bibnamefont{Rastello}}, \bibnamefont{and}
  \bibinfo{author}{\bibfnamefont{L.}~\bibnamefont{Parlato}},
  \bibinfo{journal}{IEEE Trans. Appl. Super.} \textbf{\bibinfo{volume}{9}},
  \bibinfo{pages}{3866} (\bibinfo{year}{1999}).

\bibitem[{\citenamefont{Weiser et~al.}(1981)\citenamefont{Weiser, Strom, Wolf,
  and Gubser}}]{Weiser_4888}
\bibinfo{author}{\bibfnamefont{K.}~\bibnamefont{Weiser}},
  \bibinfo{author}{\bibfnamefont{U.}~\bibnamefont{Strom}},
  \bibinfo{author}{\bibfnamefont{S.~A.} \bibnamefont{Wolf}}, \bibnamefont{and}
  \bibinfo{author}{\bibfnamefont{D.~U.} \bibnamefont{Gubser}},
  \bibinfo{journal}{J. Appl. Phys.} \textbf{\bibinfo{volume}{52}},
  \bibinfo{pages}{4888} (\bibinfo{year}{1981}).

\end{thebibliography}

\end{subequations}
\end{document}